\documentclass[prl,superscriptaddress,amsfonts,amssymb,amsmath,floats,twocolumn,aps,footinbib,shownopacs]{revtex4-2}
\usepackage{amsmath}
\usepackage{amsfonts}
\usepackage{graphicx}
\usepackage{color}
\usepackage{xcolor}
\usepackage{float}
\usepackage{hyperref}
\usepackage{soul}

\begin{document}
\title{High-quality poor man's Majorana bound states from cavity embedding}
\author{\'Alvaro G\'omez-Le\'on}
\affiliation{Institute of Fundamental Physics IFF-CSIC, Calle Serrano 113b, 28006 Madrid, Spain}
\author{Marco Schir\`o}
\affiliation{JEIP, UAR 3573 CNRS, Coll\`ege de France,   PSL  Research  University, 11,  place  Marcelin  Berthelot,75231 Paris Cedex 05, France}
\author{Olesia Dmytruk}
\affiliation{CPHT, CNRS, École polytechnique, Institut Polytechnique de Paris, 91120 Palaiseau, France}

\date{\today}

\begin{abstract}
Poor man’s Majorana Bound States (MBS) arise in minimal Kitaev chains when the parameters are fine-tuned to a sweet spot. We consider an interacting two-site Kitaev chain coupled to a single-mode cavity and show that the sweet spot condition can be controlled with the cavity frequency and the hopping between sites. 
Furthermore, we demonstrate that photon-mediated effective interactions can be used to screen intrinsic interactions, improving the original quality of the MBS. We describe experimental signatures in the cavity transmission to detect their presence and quality. Our work proposes a new way to tune poor man's MBS in a quantum dot array coupled to a cavity. 
\end{abstract}

\maketitle

\paragraph{\textbf{Introduction.}---}
The Kitaev chain is the canonical model for the appearance of Majorana bound states (MBS)~\cite{kitaev2001unpaired}. Theoretical proposals to realize the Kitaev chain Hamiltonian in an array of quantum dots connected by  superconductors were  put forth in~\cite{sau2012realizing,leijnse2012parity,fulga2013adaptive}. In a two-site Kitaev chain "poor man's MBS" emerge when the parameters are fine-tuned to a sweet spot, such that the chemical potential is tuned to zero and the hopping equals the superconducting pairing~\cite{leijnse2012parity}. In its simplest form, the coherent single particle tunneling is externally controlled by the gate voltages between dots, while the tunneling of Cooper pairs happens via virtual states in the superconductor, when electrons or holes are simultaneously created or annihilated in pairs.
Hence, in realizations involving quantum dots, tuning to the sweet spot can be achieved by controlling the tunneling with gate voltages, while the pairing will typically be fixed microscopically from the superconductor properties. Recently, a number of theoretical works~\cite{liu2022tunable,tsintzis2022creating,miles2023kitaev,samuelson2024minimal,tsintzis2024majorana,souto2023probing,liu2023enhancing,luna2024fluxtunable,liu2024coupling} offered new insights into the quantum dot-based platform and a new route towards the experimental realization of MBS~\cite{souto2024subgap}.
On the experimental side, a minimal Kitaev chain  of two-sites has been realized in a platform based on quantum dots in nanowires~\cite{dvir2023realization,zatelli2023robust} and a two-dimensional electron gas~\cite{haaf2023engineering}, while  a three-site Kitaev chain  was realized in~\cite{bordin2024signatures}. The effect of electron-electron interactions in quantum dots-based platforms is important as they can lead to the hybridization between MBS, deteriorating their quality~\cite{tsintzis2022creating,souto2024subgap}. Several markers and strategies have
been proposed to minimize these effects. For example,
on top of the degeneracy between even and odd parity ground states, a large Majorana
polarization suggests good quality MBS~\cite{tsintzis2022creating,souto2024subgap}. However, none
have found a way to completely remove the detrimental
role of many-body effects.\\
\begin{figure}[t]
    \centering
    \includegraphics[width=0.9\columnwidth]{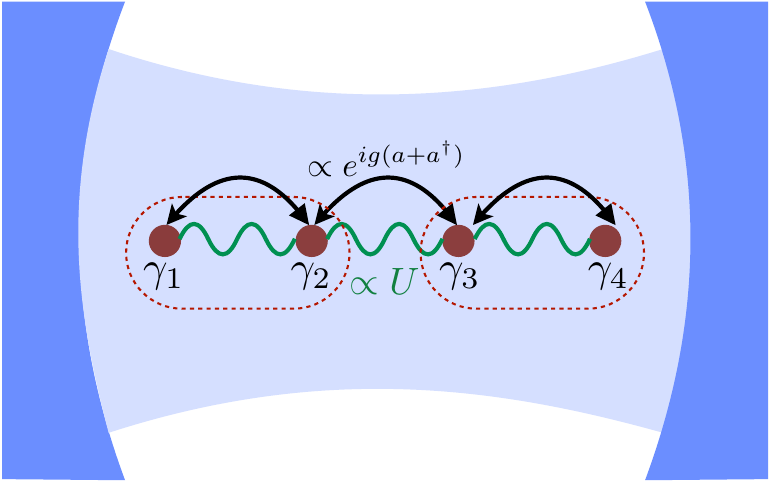}
    \caption{The setup consists of a two-site Kitaev chain described by the Majorana operators $\gamma_1 - \gamma_4$ (brown circles) coupled to a single mode cavity with frequency $\omega_c$ through a Peierls phase $\propto e^{ig(a+a^\dag)}$ (black arrows). The strength of the light-matter coupling is $g$. The on-site chemical potential is $\mu$, the hopping amplitude between two sites is $t$, the $p$-wave pairing is $\Delta$, and the strength of electron interaction is $U$ (green lines).
    }
    \label{fig:Scheme}
\end{figure}

In this Letter, we show that coupling a minimal interacting Kitaev chain to an off resonant cavity mode offers a new and largely tunable platform to control poor man's MBS. Cavity embedding has been recently introduced as a novel way for probing and controlling quantum matter~\cite{garciavidal2021manipulating,schlawin2022cavity}, with applications ranging from topological materials~\cite{ciuti2021cavity,appugliese2022breakdown,perez2022topology,dmytruk2022controlling,nguyen2023electron,pérezgonzález2023lightmatter,vlasiuk2023cavity,shaffer2024entanglement,nguyen2024electron} to strongly correlated systems~\cite{jarcNature2023,kozin2024cavity} and mesoscopic systems ~\cite{delbecq2011coupling,petersson2012coupling,frey2012dipole,basset2013single,delbecq2013engineering,viennot2014out,viennot2015coheret,stockklauser2015microwave,mi2016strong,cottet2017cavity,mi2018electron,samkharadze2018strong,landig2018electron}, including those hosting MBS ~\cite{trif2012resonantly,cottet2013squeezing,dmytruk2015cavity,dmytruk2016josephson,dartiailh2017direct,trif2019braiding,contamin2021topological,dmytruk2023microwave,bacciconi2024topological,dmytruk2023hybrid}. In the microwave regime the cavity is typically  implemented by a superconducting resonator~\cite{Blais2004}, whose frequency~\cite{Palacios-Laloy2008} or number of photons can be tuned. 
Having this setup in mind we consider a model for a double quantum dot (DQD), realising a minimal two-site interacting Kitaev chain, coupled to a single mode cavity (see Fig.~\ref{fig:Scheme}). In the large detuning regime, when the cavity is off resonant with respect to the DQD transitions, we can adiabatically eliminate the photons and obtain
an effective Hamiltonian for the interacting Kitaev chain. We show that the cavity photons renormalize the chemical potential, the hopping and the interaction, leading to MBS that can be externally controlled.
We demonstrate that the condition for the fine-tuned sweetspot is modified in the presence of the cavity. In particular, that by tuning the cavity frequency and the DQD hopping it is possible to obtain isolated MBS for a wide range of light-matter couplings and superconducting pairings. Furthermore, we find that interdots interactions, that are always detrimental for the existence of MBS, can be screened by photon mediated interactions, enhancing the localization properties and overall quality of the original MBS.
Our results perfectly agree with exact diagonalization results for the full many-body spectrum.
 
\paragraph{\textbf{Two-site Kitaev chain coupled to cavity.}---}
We consider an interacting, two-site Kitaev chain coupled to a single mode cavity
\begin{align}
H =& -\mu\left(c_{1}^{\dagger}c_{1}+c_{2}^{\dagger}c_{2}\right)-t\left(e^{-ig\left(a^{\dagger}+a\right)}c_{1}^{\dagger}c_{2} + \text{h.c.}\right)\notag\\
&+\Delta \left(c_{1}c_{2}+\text{h.c.}\right) +\omega_c a^\dag a+U c_1^\dagger c_1 c_2^\dagger c_2,
\label{eq:N=2HKitaevCavity}
\end{align}
where $c_{1,2}^\dag$ ($c_{1,2}$) are fermionic creation (annihilation) operators at site $1,2$ and $a^\dag $ ($a$) are photonic creation (annihilation) operators. Here,  $\mu$ is the chemical potential, $t$ is the hopping,  $\Delta$ is a $p$-wave superconducting pairing potential, $\omega_c$ is a cavity frequency, $U$ is the inter-dot interaction, and $g$ is the light-matter coupling that we write in the form of a Peierls phase modulating the inter-dot hopping.
This model arises as an effective description of a DQD architecture interfaced with a $s$-wave superconductor~\cite{sau2012realizing,leijnse2012parity}, integrated in a microwave resonator. 

\paragraph{\textbf{Isolated chain.}---}
The isolated two-site Kitaev chain Hamiltonian can be exactly diagonalized in the many-body basis:
$\{ |0_1 0_2 \rangle  , |1_10_2  \rangle  ,|0_11_2  \rangle  , |1_11_2  \rangle  \}$ $ = $ $\{ |0_{1} 0_{2} \rangle  , c^\dag_{1}|0_{1} 0_{2} \rangle , c^\dag_{2}|0_{1} 0_{2} \rangle , c^\dag_{1}c^\dag_{2}|0_{1} 0_{2} \rangle \}$, with $| 0_10_2 \rangle$ being the state with no electrons. 
The pairing term, proportional to $\Delta$, mixes particles and holes, making the total number of particles no longer a conserved quantity. Instead, the state parity is conserved and allows us to separate the electronic part of the Hamiltonian into even $\{ |0_{1} 0_{2} \rangle  , |1_{1} 1_{2} \rangle \}$ and odd parity sectors $\{  |1_{1} 0_{2} \rangle ,|0_{1} 1_{2} \rangle  \}$, with energies $E_\pm^{\text{even}} = U/2-\mu \pm \sqrt{\Delta^2+\left( U/2-\mu \right)^2} $ and $E_\pm^\text{odd} = -\mu \pm t$, respectively.

If we express the isolated chain Hamiltonian in the Majorana basis, $c_{j}=\frac{1}{2}\left(\gamma_{2j-1}+i\gamma_{2j}\right)$ with $\left\{ \gamma_{j},\gamma_{k}\right\} =2\delta_{j,k}$,
\begin{align}
    H  \underset{g\to0}{=} & \frac{U}{4}-\mu-\frac{i}{2}\left(\mu-\frac{U}{2}\right)\left(\gamma_{1}\gamma_{2}+\gamma_{3}\gamma_{4}\right) \label{eq:MajoranaH0}\\
    &+i\frac{\Delta-t}{2}\gamma_{1}\gamma_{4}+i\frac{\Delta+t}{2}\gamma_{2}\gamma_{3}-\frac{U}{4}\gamma_{1}\gamma_{2}\gamma_{3}\gamma_{4} , \nonumber
\end{align}
it can be seen that at the sweet spot $t=\Delta$ and $\mu=U/2$, isolated MBS $\gamma_{1,4}$ will only emerge for the case $U=0$. Otherwise, the last term hybridizes their wave functions and spoils their properties~\cite{leijnse2012parity}.
Experimentally, this strict condition requires fine tuning and to completely screen particle interactions, although recent works~\cite{tsintzis2022creating} have shown that in the presence of interactions, one can simultaneously require a ground state degeneracy and a large Majorana polarization~\cite{Aksenov2020MP} in order to still create good quality MBS.\\
\textit{In this work we demonstrate that coupling the chain to a cavity allows to control the sweet spot condition required to find MBS, and that even for the case of $U\neq0$, one can use photon-mediated interactions to effectively screen the interactions between the QDs, giving rise to isolated MBS.}
\paragraph{\textbf{Effective Hamiltonian.}---}
We write an effective matter Hamiltonian for the subspace with $n$ photons using the projectors method~\cite{FESHBACH1958357}, which to lowest order reads (details in the SM):
\begin{equation}
    \tilde{H} (n,\tau) = PH_{0}P-i\int_{0}^{\tau}d\tau^\prime PH_{1}e^{-iH_{0}\tau^\prime}H_{1}e^{iH_{0}\tau^\prime}P, \label{eq:EffectiveH1}
\end{equation}
where $P$ is the projector onto the $n$-photons Fock subspace, $Q=1-P$ is the projector onto the $n\pm 1$-photons subspace, $H_0$ is the diagonal part of the Hamiltonian~(\ref{eq:N=2HKitaevCavity}) in each Fock subspace and $H_1$ describes photon transitions between different Fock subspaces.
Importantly, as in the presence of a cavity Eq.~\eqref{eq:N=2HKitaevCavity} shows that the DQD couples to photons via the hopping term only~\cite{kozin2023quantum,dmytruk2023hybrid}, we can simplify the analysis and restrict the study to the odd subspace.
The projection onto a Fock subspace with a well-defined number of photons is justified in the large detuning regime $\omega_c\gg t,\Delta$, where photon transitions are very unlikely and their effect can be efficiently encoded via adiabatic elimination to lowest order. Interestingly, this leads to an $n$-dependent expression, indicating that the effective Hamiltonian depends on the number of photons in which the cavity is initially prepared.
Also, Eq.~\eqref{eq:EffectiveH1} generally is a time-dependent effective Hamiltonian, but for the large detuning regime one can perform a Rotating Wave Approximation (RWA) and ignore the time-dependent terms, which approximately average to zero at relevant time-scales. 
In this regime the effective Hamiltonian reads:
\begin{align}
    \tilde{H}\left(n\right) =& \tilde{U}\left(n\right)c_{1}^{\dagger}c_{1}c_{2}^{\dagger}c_{2}- \tilde{\mu}\left(n\right)(c_{1}^{\dagger}c_{1}+c_{2}^{\dagger}c_{2}) \nonumber \\
    &+\Delta(c_{1}c_{2}+c_{2}^{\dagger}c_{1}^{\dagger})- \tilde{t}\left(n\right)(c_{1}^{\dagger}c_{2}+c_{2}^{\dagger}c_{1}) . \label{eq:Heff}
\end{align}
Eq.~\eqref{eq:Heff} shows that the cavity affects the chemical potential, the hopping and the interaction between particles. In particular, their expressions are:
\begin{align}
    \tilde{U}\left(n\right) =& U-2n\omega_{c}+\frac{2\left(n+1\right)\omega_{c}\tilde{\kappa}_{n+1}^{2}}{\omega_{c}^{2}-\tilde{\omega}_{n+1}^{2}}-\frac{2n\omega_{c}\tilde{\kappa}_{n}^{2}}{\omega_{c}^{2}-\tilde{\omega}_{n}^{2}} \label{eq:EffU1} ,\\
    \tilde{\mu}\left(n\right) =& \mu-n\omega_{c}+\frac{\left(n+1\right)\omega_{c}\tilde{\kappa}_{n+1}^{2}}{\omega_{c}^{2}-\tilde{\omega}_{n+1}^{2}}-\frac{n\omega_{c}\tilde{\kappa}_{n}^{2}}{\omega_{c}^{2}-\tilde{\omega}_{n}^{2}} ,\\
    \tilde{t}\left(n\right) =& te^{-\frac{g^{2}}{2}}L_{n}(g^{2})-\frac{(n+1)\tilde{\omega}_{n+1}\tilde{\kappa}_{n+1}^{2}}{\omega_{c}^{2}-\tilde{\omega}_{n+1}^{2}}-\frac{n\tilde{\omega}_{n}\tilde{\kappa}_{n}^{2}}{\omega_{c}^{2}-\tilde{\omega}_{n}^{2}}\label{eq:tEff} ,
\end{align}
where $\tilde{\kappa}_{n} =gte^{-\frac{g^{2}}{2}}\,_{1}F_{1}\left(1-n;2;g^{2}\right)$ and $\tilde{\omega}_{n} = te^{-\frac{g^{2}}{2}}\left[L_{n-1}\left(g^{2}\right)+L_{n}\left(g^{2}\right)\right]$.
Also, $L_n(g^2)$ is the $n$th Laguerre polynomial and $\,_{1}F_{1}(-n;2;g^{2})$ the Kummer confluent hypergeometric function. Eqs.~\eqref{eq:EffU1} to~\eqref{eq:tEff} show that the original Coulomb interaction, chemical potential and hopping can be externally tuned by changing the light-matter interaction $g$, the cavity frequency $\omega_c$ and the cavity state preparation $n$. Again, it is also useful to express Eq.~\eqref{eq:Heff} in terms of Majorana operators:
\begin{align}
    \tilde{H}\left(n\right)=& \frac{\tilde{U}\left(n\right)}{4}-\tilde{\mu}\left(n\right)-\frac{\tilde{U}\left(n\right)}{4}\gamma_{1}\gamma_{2}\gamma_{3}\gamma_{4} \nonumber \\
    &-\frac{i}{2}\left[\tilde{\mu}\left(n\right)-\frac{\tilde{U}\left(n\right)}{2}\right]\left(\gamma_{1}\gamma_{2}+\gamma_{3}\gamma_{4}\right) \nonumber \\
    &+i\frac{\Delta-\tilde{t}\left(n\right)}{2}\gamma_{1}\gamma_{4}+i\frac{\Delta+\tilde{t}\left(n\right)}{2}\gamma_{2}\gamma_{3} .
    \label{eq:HeffMajoranaBasis}
\end{align}
It shows that in order to have isolated MBS one must simultaneously fulfill the three conditions: $\tilde{\mu}(n) = \tilde{U}(n)/2$, $\Delta=\tilde{t}(n)$ and $\tilde{U}(n)=0$. The first condition leads to $\mu=U/2$, which fixes the chemical potential in terms of the inter-dot interaction. The second and third determine the values of the hopping $t$ and cavity frequency $\omega_c$, required to isolate MBS in terms of the given microscopic values of light-matter coupling $g$, pairing $\Delta$, and interaction $U$.\\
\textit{Crucially, these MBS effectively are analogous to those of the isolated chain due to the large detuning between the cavity and the system, which efficiently suppresses hybridization with the cavity mode. 
On top of this, the intrinsic interactions can now be completely screened, improving the quality of the original MBS.}
\paragraph{\textbf{Results.}---}
The simplest scenario involves a cavity in its ground state ($n=0$) coupled to the interacting, minimal Kitaev chain.
The photon-mediated interaction $\tilde{U}(n=0)$ from Eq.~\eqref{eq:EffU1} takes a particularly simple form, and the condition to screen the interactions $\tilde{U}(n=0)=0$ reads:
\begin{equation}
    U= -\frac{2\omega_{c}g^{2}t^{2}e^{-g^{2}}}{\omega_{c}^{2}-t^{2}e^{-g^{2}}\left(2-g^{2}\right)^{2}} . \label{eq:EffU0}
\end{equation}
In addition, Eq.~\eqref{eq:tEff} requires $\tilde{t}(n=0) = \Delta$, or equivalently:
\begin{equation}
\Delta = t e^{-\frac{g^2}{2}}  -\frac{g^2 t^3 \left(2-g^2\right) e^{-3 g^2/2}  }{\omega_c ^2- t^2e^{-g^2} \left(2-g^2\right)^2 }, \label{eq:tEff2}
\end{equation}
in order to find isolated MBSs. We solve Eqs.~\eqref{eq:EffU0} and \eqref{eq:tEff2} for fixed microscopic parameters $U$, $g$ and $\Delta$, to find the necessary values for the hopping and the cavity frequency $(t,\omega_c)$, that drive the system to the sweet spot.
Importantly, the rhs of Eq.~\eqref{eq:EffU0} for large detuning can only take negative values, which indicates that for the cavity in its ground state, it is only possible to screen attractive interactions or equivalently, that the photonic vacuum only mediates repulsive interactions between MBSs. This case is analyzed in detail in the SM.


We now consider a more relevant situation for the realization of minimal Kitaev chains with interacting DQDs, which is the case of repulsive interactions.
Screening repulsive interactions for a largely detuned cavity requires to prepare the cavity with an average number of photons $n>0$.
In that case, the second term in Eq.~\eqref{eq:EffU1} makes the photon-mediated interaction attractive, such that it can compensate for intrinsic repulsive interactions between particles $U>0$.
This can be done by driving the resonator until the average value of photons in the isolated cavity is approximately one. In this large detuning regime the cavity will remain in this meta-stable state until photon losses drive it to its ground state.
We assume a cavity prepared in the Fock subspace with $n=1$ and numerically find the sweet spot values $(t,\omega_c)$ from the conditions to find isolated MBS: $\tilde{U}(n=1)=0$ and $\tilde{t}(n=1)=\Delta$, with $\mu=U/2$.
\begin{figure}
    \centering
    \includegraphics[width=0.9\columnwidth]{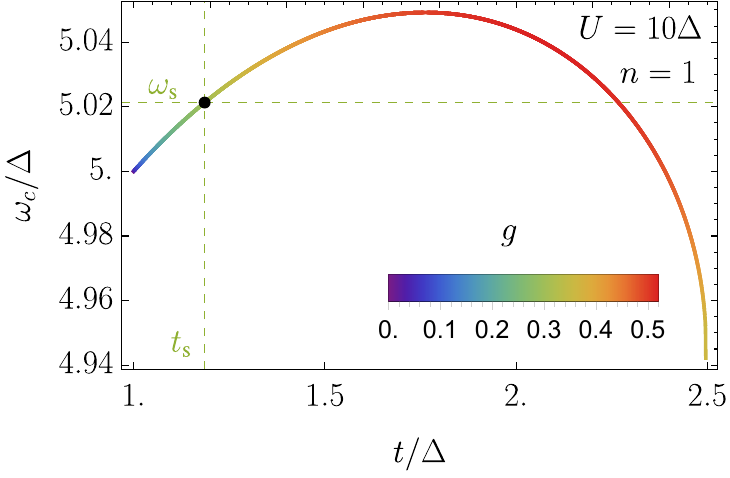}
    \caption{Sweet spot values in an interacting minimal Kitaev chain. The color of the curve indicates the value of $g$. The cavity is prepared in a $n=1$ photon state and the other parameters are $\mu=U/2$ and $U = 10\Delta$.}
    \label{fig:SweetSpot1}
\end{figure}
In Fig.~\ref{fig:SweetSpot1} we show the region of existence of these solutions for different values of the hopping and the cavity frequency.
In the absence of the cavity the sweet spot is just a vertical line at $t=\Delta$, only valid for the case $U=0$. Instead, the coupling to the cavity photons transforms the sweet spot into a curve where interactions between MBS are completely screened.
Note that the color code shows that for a wide variety of microscopic light-matter couplings, the sweet spot can be reached by adjusting the DQD hopping and the cavity frequency.

\begin{figure}[t]
    \centering
    \includegraphics[width=0.9\columnwidth]{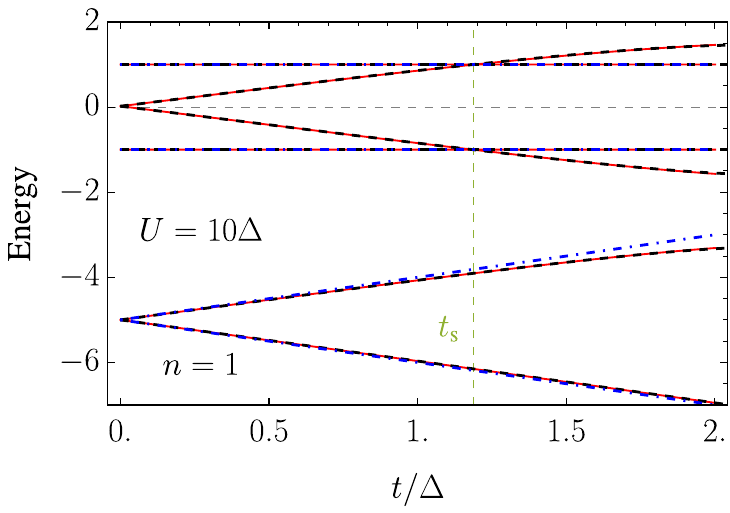}
    \caption{Spectrum of the interacting two-site Kitaev chain coupled to a cavity with one photon $n=1$. The blue dot-dashed line corresponds to the isolated case ($g$ = 0) given by $E_\pm^{\text{odd}}$ and $E_\pm^{\text{even}}$. Red solid and black dashed lines correspond to exact diagonalization for $N_{max}=20$ for the case $g = 0.3$ and its analytical  approximation with Eq.~\eqref{eq:Heff}, respectively. The green dashed vertical line indicates the sweet spot hopping. The other parameters are $\omega_c=5.02\Delta$, $U=10\Delta$ and $\mu=U/2$.
    }
    \label{fig:Spectrum1}
\end{figure}

Fig.~\ref{fig:Spectrum1}, shows the spectrum for a particular case indicated with a black dot in Fig.~\ref{fig:SweetSpot1}.
For the isolated case $g=0$ (blue dot-dashed),
the condition $\mu=U/2$ (equivalent to require a maximum Majorana polarization), pushes the even subspace levels $E_\pm^{\text{even}}$ high in energy due to the large repulsive interaction.
Hence, in this case it is not possible to find the degeneracy between the even and the odd ground states of the sweet spot by just tuning the hopping.
This changes for $g\neq0$, as the preparation of the cavity in the one-photon state and the subsequent interaction with the Kitaev chain produces a copy of the odd-subspace energy band close to zero energy (red solid line). It is now possible to tune the hopping to the sweet spot $t_\text{s}$ (green dashed vertical line) and find an exact degeneracy between even and odd subspaces. Crucially, if the system is correctly tuned to the sweet spot (i.e. $\omega_c=\omega_s$ and $t=t_s$), the degeneracy occurs simultaneously for the excited states, indicating a symmetry around zero energy, as expected for isolated MBS. In this configuration, the resulting MBSs are of high quality and display maximum Majorana polarization.
At large $t/\Delta$ the analytical approximation (black dashed line) deviates from exact diagonalization because the bandwidth approaches the cavity frequency and the system abandons the large detuning regime.
It is important to stress that although the cavity is prepared in an excited state with one photon, in the large detuning regime the system will be in a long-lived meta-stable state only decaying via photon losses. This is because the large detuning makes the resulting many-body state close, but not identical, to a product state of the cavity and the Kitaev chain. This is crucial, because the hybridization is small enough to adiabatically eliminate the photons from the effective Hamiltonian and not change the number of photons initially prepared in the cavity, but also to allow the existence of photon-mediated interactions between particles that can be used to screen the ones intrinsic to the system. Therefore, for time-scales shorter than the photon losses, we can consider the MBSs as identical to those present in isolated Kitaev chains.

\begin{figure}[t]
    \centering
    \includegraphics[width=0.9\columnwidth]{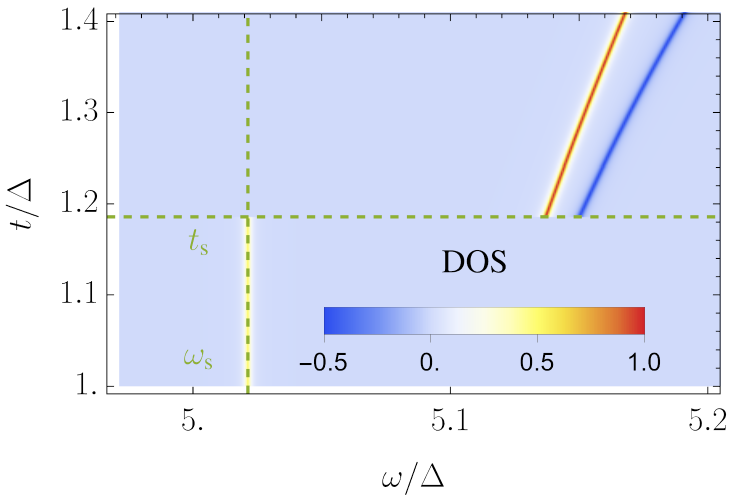}
    \caption{Density of states of the photon propagator from exact diagonalization, as a function of $\omega$ and $t$. The horizontal dashed line shows the sweet spot hopping predicted from the effective Hamiltonian. The vertical dashed line indicates the position of the bare resonator frequency $\omega_c$.}
    \label{fig:Transmission1}
\end{figure}

\paragraph{\textbf{Detection.}---}
A relevant feature of c-QED systems is that properties of the setup interacting with the cavity can be inferred from the photons themselves~\cite{Blais2004,Zueco2009}.
The photon propagator is proportional to the cavity transmission~\cite{perez2022topology} and can be measured in the transmission line of standard quantum optics experiments using homodyne detection.
We calculate the photon propagator $D(t)\equiv-i\theta(t)\langle[a(t),a^\dagger]\rangle$ to detect the presence of MBS in the chain and find that the cavity frequency shift strongly depends on the parity of the ground state. In particular, as the even subspace does not couple to the cavity, the photon propagator is unaffected when the Kitaev chain is in its ground state. Instead, when the Kitaev chain is in the odd ground state the cavity frequency shifts from $\omega_c$.
This is shown in Fig.~\ref{fig:Transmission1} for a Kitaev chain with repulsive interactions and a cavity initially prepared with one photon.
For small hopping $t/\Delta$, the ground state of the system is even and the cavity frequency is not affected by the presence of the Kitaev chain. Hence, the density of states $\text{DOS}=-\text{Im}D(\omega)/\pi$ displays a single resonance peak at the bare resonator frequency $\omega_c$ (vertical dashed line). At the sweet spot $t_\text{s}$ (horizontal dashed line), the ground state becomes odd and the DOS displays two resonance peaks with opposite sign.
This feature can be reproduced by considering the effective Hamiltonian for the odd subspace to lowest order. For a cavity with $n$ photons the photon propagator reads $D_{n}\left(\omega\right)=(n+1)/\left(\omega-\omega_{n+1}\right)-n/\left(\omega-\omega_{n}\right)$, with $\omega_{n}\equiv\omega_c+te^{-\frac{g^{2}}{2}}\left[L_{n-1}\left(g^{2}\right)-L_{n}\left(g^{2}\right)\right]$ (for more details see the SM). Therefore, for a cavity in the vacuum only one resonance is present, but for $n>0$, two peaks with opposite sign emerge.
Notice that transmission spectroscopy of the cavity allows us to find the degeneracy point between ground states, but it does not confirm that the interactions between MBS are completely screened.
That is, the cavity frequency shift predicts a change in the parity of the ground state, but not necessarily that the frequency of the resonator is exactly the one of the sweet spot. To determine its value from the experimental measurement one can also study the distance between the two peaks in the odd parity ground state, which is given by $\Delta\omega_n=\omega_{n+1}-\omega_{n}$ and it is a polynomial only on $g$. Another alternative would be to perform resonance experiments to extract the Rabi splitting, although here one must take into account the nonlinear nature of the system. Other possibilities would include non-local measurements of the MBS~\cite{GomezRuiz2018,Mendez-Cordoba2023}.

\paragraph{\textbf{Conclusions.}---}
We have studied an interacting minimal Kitaev chain coupled to a single mode cavity as a platform to engineer isolated MBS.
We have shown that in the large detuning regime one can adiabatically eliminate the cavity and produce an effective Hamiltonian for the minimal Kitaev chain where the chemical potential, hopping and inter-dot interaction can be externally controlled. This produces a tunable sweet spot for the appearance of MBS that crucially, due to photon-mediated effective interactions, can completely screen the detrimental inter-dot interactions that are ubiquitous is these systems and hybridize the MBS.
In particular we have demonstrated that ground state degeneracy and maximum Majorana polarization are necessary, but insufficient conditions to completely isolate the MBS when interactions are present. An additional condition must be imposed and can only be fulfilled when photon mediated interactions are present.\\
Our work paves the way for the creation and detection of MBS in minimal Kitaev chains and highligths the important role of cavity-mediated interactions.
In the future it would be interesting to study higher order photon correlations for the characterization and the presence of MBS or to extend this analysis to larger Kitaev chains.
\paragraph{\textbf{Acknowledgments.}---}
MS acknowledges financial support from the ERC consolidator grant No.~101002955 - CONQUER. OD acknowledges support from the ERC Starting Grant, Grant Agreement No.~101116525. AGL acknowledges support from the MOLAR QuantERA project. 

\newpage

\bibliography{bibliographyDraftKitaev}

\clearpage

\widetext
\pagebreak
\setcounter{equation}{0}
\setcounter{figure}{0}
\setcounter{table}{0}
\setcounter{page}{1}
\renewcommand{\theequation}{S\arabic{equation}}
\setcounter{figure}{0}
\renewcommand{\thefigure}{S\arabic{figure}}
\renewcommand{\thepage}{S\arabic{page}}
\renewcommand{\thesection}{S\arabic{section}}
\renewcommand{\thetable}{S\arabic{table}}
\makeatletter

\renewcommand{\thesection}{\arabic{section}}
\renewcommand{\thesubsection}{\thesection.\arabic{subsection}}
\renewcommand{\thesubsubsection}{\thesubsection.\arabic{subsubsection}}

\renewcommand{\bibnumfmt}[1]{[S#1]}

\renewcommand{\citenumfont}[1]{S#1}

\begin{widetext}
	
	\newpage
	\onecolumngrid
	\bigskip 
	
	\begin{center}
		\large{\bf Supplemental Material to `High-quality poor man's Majorana bound states from cavity embedding' \\}
\author{\'Alvaro G\'omez-Le\'on}
\affiliation{Institute of Fundamental Physics IFF-CSIC, Calle Serrano 113b, 28006 Madrid, Spain}
\author{Marco Schir\`o}
\affiliation{JEIP, UAR 3573 CNRS, Coll\`ege de France,   PSL  Research  University, 11,  place  Marcelin  Berthelot,75231 Paris Cedex 05, France}
\author{Olesia Dmytruk}
\affiliation{CPHT, CNRS, École polytechnique, Institut Polytechnique de Paris, 91120 Palaiseau, France}

	\end{center}

\section{Derivation of the effective Hamiltonian}
An effective description of the 2-sites Kitaev chain coupled to a cavity, in terms of the matter degrees of freedom only, can be obtained by means of the projectors method~\cite{FESHBACH1958357}.
To lowest order, the effective Hamiltonian in the relevant subspace described by a projector $P$, once the irrelevant subspace described by the projector $Q$ (with $P+Q=1$) is adiabatically eliminated, can be expressed as:
\begin{equation}
    \tilde{H}\left(n,\tau\right)=PH_{0}P-i\int_{0}^{\tau}d\tau^\prime PH_{1}e^{-iH_{0}\tau^\prime}H_{1}e^{iH_{0}\tau^\prime}P
\end{equation}
where $H_0$ corresponds to the unperturbed part of the Hamiltonian and $H_1 = H-H_0$ to the perturbation terms. Importantly, for this expression to be valid we must separate $H_0$ and $H_1$ such that they fulfill $PH_0Q=0=QH_0P$, $PH_1Q=PH_1$ and $QH_1P=H_1P$.
Note that the effective Hamiltonian will generally be time-dependent, because the effect of the subspace $Q$ might be complex, and that the description will fail if the perturbation resonantly couples levels of the unperturbed Hamiltonian (can be shown by working in the eigenstates basis of $H_0$). Intuitively this is due to the fact that resonant levels actively contribute to the dynamics, so it is not possible to obtain the effective dynamics of the subspace $P$.

In our particular case we consider the full Hamiltonian $H=H_0+H_1$ and separate the dominant and perturbative parts as:
\begin{equation}
    H_{0}=\omega_c\sum_{n=0}^{\infty}nY^{n,n}-\mu\left(c_{1}^{\dagger}c_{1}+c_{2}^{\dagger}c_{2}\right)+\Delta \left(c_{1}c_{2} + c_{2}^{\dagger}c_{1}^{\dagger} \right)+U c_{1}^{\dagger}c_{1}c_{2}^{\dagger}c_{2}-t\sum_{n=0}^{\infty}e^{-\frac{g^{2}}{2}}L_{n}\left(g^{2}\right)\left(c_{2}^{\dagger}c_{1}+c_{1}^{\dagger}c_{2}\right)Y^{n,n} \label{eq:DiagH}
\end{equation}
and
\begin{equation}
    H_{\text{1}}=-\sum_{n,m\neq n}^{\infty}\left[t_{m,n}\left(g\right)c_{2}^{\dagger}c_{1}+t_{m,n}\left(-g\right)c_{1}^{\dagger}c_{2}\right]Y^{m,n},
\end{equation}
where we have expressed the total Hamiltonian in the basis of number of photon states $|n\rangle$ and defined their corresponding bosonic Hubbard operators $Y^{n,m} \equiv |n \rangle \langle m|$~\cite{perez2023light}.
There, the definition of the photon-dependent hopping is
\begin{equation}
    t_{m,n}(g) = t e^{-\frac{g^{2}}{2}}\left(ig\right)^{n-m}\sqrt{\frac{n!}{m!}}\,_{1}\tilde{F}_{1}\left(-m;n+1-m;g^{2}\right) ,
\end{equation}
with $\,_{1}\tilde{F}_{1}\left(a;b;z\right)$ is the regularized confluent hypergeometric function. The diagonal components are proportional to the Laguerre polynomials:
\begin{equation}
    t_{n,n}(g) = t e^{-\frac{g^2}{2}} L_n(g^2)
\end{equation}
Therefore, $H_0$ contains the diagonal Fock states elements, which include a renormalization of the hopping due to virtual photon absorption/emission processes, and $H_1$ the real photon transitions $m\neq n$.
Importantly, note that the even subspace $|0_1,0_2\rangle$, $|1_1,1_2\rangle$ is decoupled from the cavity field. This allows us to work with the Hamiltonian for the odd subspace only. However, the Hamiltonian for the odd subspace still is infinite dimensional due to the coupling to the cavity.
For our purposes, it will be enough to just consider the set of coupled Fock subspaces $n$ and $n\pm 1$, and estimate the effective Hamiltonian in the subspace $n$, once the $n\pm 1$ subspaces have been adiabatically eliminated.
Hence, in our case the projectors are particularized to $P=|n \rangle\langle n|$ and $Q=|n-1 \rangle\langle n-1|+|n+1 \rangle\langle n+1|$.

The calculation of the effective Hamiltonian in the fermionic odd-subspace for a cavity with $n$ photons results in the following expression:
\begin{align}
    \tilde{H}_{\text{odd}}\left(n,\tau\right)=& \left(n\omega_{c}-\mu\right)\sigma_{0}-te^{-\frac{g^{2}}{2}}L_{n}\left(g^{2}\right)\sigma_{x} -\frac{\left(n+1\right)\tilde{\kappa}_{n+1}^{2}}{\omega_{c}^{2}-\tilde{\omega}_{n+1}^{2}}\left(\omega_{c}\sigma_{0}-\tilde{\omega}_{n+1}\sigma_{x}\right)+\frac{n\tilde{\kappa}_{n}^{2}}{\omega_{c}^{2}-\tilde{\omega}_{n}^{2}}\left(\omega_{c}\sigma_{0}+\tilde{\omega}_{n}\sigma_{x}\right) \nonumber \\
    &+\frac{\left(n+1\right)\tilde{\kappa}_{n+1}^{2}e^{-i\omega_{c}\tau}}{\omega_{c}^{2}-\tilde{\omega}_{n+1}^{2}}\left[\omega_{c}\cos\left(\tilde{\omega}_{n+1}\tau\right)+i\tilde{\omega}_{n+1}\sin\left(\tilde{\omega}_{n+1}\tau\right)\right]\sigma_{0} \nonumber \\
    &-\frac{\left(n+1\right)\tilde{\kappa}_{n+1}^{2}e^{-i\omega_{c}\tau}}{\omega_{c}^{2}-\tilde{\omega}_{n+1}^{2}}\left[\tilde{\omega}_{n}\cos\left(\tilde{\omega}_{n+1}\tau\right)+i\omega_{c}\sin\left(\tilde{\omega}_{n+1}\tau\right)\right]\sigma_{x} \nonumber \\
    &-\frac{n\tilde{\kappa}_{n}^{2}e^{i\omega_{c}\tau}}{\omega_{c}^{2}-\tilde{\omega}_{n}^{2}}\left[\omega_{c}\cos\left(\tilde{\omega}_{n}\tau\right)-i\tilde{\omega}_{n}\sin\left(\tilde{\omega}_{n}\tau\right)\right]\sigma_{0} -\frac{n\tilde{\kappa}_{n}^{2}e^{i\omega_{c}\tau}}{\omega_{c}^{2}-\tilde{\omega}_{n}^{2}}\left[\tilde{\omega}_{n}\cos\left(\tilde{\omega}_{n}\tau\right)-i\omega_{c}\sin\left(\tilde{\omega}_{n}\tau\right)\right]\sigma_{x} \label{eq:EffH1}
\end{align}
where we have introduced the frequency $\tilde{\omega}_{n}\equiv te^{-\frac{g^{2}}{2}}\left[L_{n-1}\left(g^{2}\right)+L_{n}\left(g^{2}\right)\right]$, $\tilde{\kappa}_n \equiv g t e^{-\frac{g^{2}}{2}}\,_{1}F_{1}\left(1-n;2;g^{2}\right)$ and the Pauli matrices are the standard ones acting on the odd subspace with basis elements $|1_1,0_2\rangle$ and $|0_1,1_2\rangle$.
The first line of Eq.~\eqref{eq:EffH1} contains the time-independent part, with the last two terms being the correction to $H_0$. The rest contains the time-dependent contributions, which can be neglected if $\omega_c \gg t$. In that case, a RWA indicates that the rapidly oscillating terms, proportional to $e^{\pm i\omega_c\tau}$, average to zero and can be safely discarded.

In order to extract the effective interactions that the effective Hamiltonian encodes, we rewrite the effective Hamiltonian in terms of fermionic operators.
From the even subspace we have the original Hamiltonian, because it is not affected by the cavity (remember that we defined $|1_1,1_2\rangle=c_1^\dagger c_2^\dagger|0_1,0_2\rangle$:
\begin{equation}
    \tilde{H}_\text{even} = \left(U-2\mu\right)|2\rangle\langle2|-\Delta (|0\rangle \langle 2|+|2\rangle \langle 0|)= \left(U-2\mu\right)c_{1}^{\dagger}c_{1}c_{2}^{\dagger}c_{2}+\Delta\left(c_{1}c_{2}+c_{2}^{\dagger}c_{1}^{\dagger}\right)
\end{equation}
Instead, for the odd subspace we find:
\begin{align}
    \tilde{H}_\text{odd}\left(n,\tau\right)=&h_{\text{O}}\left(n,\tau\right)\left(|1,0\rangle\langle0,1|+|0,1\rangle\langle1,0|\right)+h_{\text{D}}\left(n,\tau\right)\left(|1,0\rangle\langle1,0|+|0,1\rangle\langle0,1|\right) \nonumber \\
    =&h_{\text{O}}(n,\tau)\left(c_{1}^{\dagger}c_{2}+c_{2}^{\dagger}c_{1}\right)+h_{\text{D}}\left(n,\tau\right)\left(c_{1}^{\dagger}c_{1}+c_{2}^{\dagger}c_{2}-2c_{1}^{\dagger}c_{2}^{\dagger}c_{2}c_{1}\right)
\end{align}
where we have defined $h_{\text{D}}(n,\tau)$ as the time-dependent coefficients that multiply $\sigma_0$ in Eq.~\eqref{eq:EffH1} and $h_{\text{O}}(n,\tau)$ as the ones that multiply $\sigma_x$.
Note that the shift in the local energies of the odd-subspace produces an energy difference with the local energies in the even subspace, and then leads to an effective, time-dependent interaction term $-2h_{\text{D}}(n,\tau)c_{1}^{\dagger}c_{1}c_{2}^{\dagger}c_{2}$.
Therefore, the effective Hamiltonian for the $n$-photons subspace after adiabatic elimination of the $n\pm1$ subspaces reads:
\begin{equation}
    \tilde{H}\left(n,\tau\right)=\left[U-2\mu-2h_{\text{D}}\left(n,\tau\right)\right]c_{1}^{\dagger}c_{2}^{\dagger}c_{2}c_{1}+\Delta\left(c_{1}c_{2}+c_{2}^{\dagger}c_{1}^{\dagger}\right)+h_{\text{D}}\left(n,\tau\right)\left(c_{1}^{\dagger}c_{1}+c_{2}^{\dagger}c_{2}\right)+h_{\text{O}}\left(n,\tau\right)\left(c_{1}^{\dagger}c_{2}+c_{2}^{\dagger}c_{1}\right) \label{eq:EffHam2}
\end{equation}
Note that it has a time-dependent correction to the interaction term $ U-2\mu-2h_{\text{D}}\left(n,\tau\right) \equiv \tilde{U}\left(n,\tau\right)$, the local energies $ -h_{\text{D}}\left(n,\tau\right) \equiv \tilde{\mu}$ and the hopping $-h_{\text{O}}\left(n,\tau\right) \equiv \tilde{t}$.
\section{Large detuning regime}
The effective Hamiltonian from Eq.~\eqref{eq:EffHam2} highly simplifies if the cavity is largely detuned to higher frequencies, because photon transitions between the $n$ and $n+1$ subspaces are very unlikely.
In that scenario, the use of the RWA allows to neglect the time-dependent terms because they approximately average to zero at relevant time-scales.
Hence, we can simplify the effective Hamiltonian for the odd subspace to just the first line of Eq.~\eqref{eq:EffH1}
\begin{equation}
    \tilde{H}\left(n\right) = \tilde{U}\left(n\right)c_{1}^{\dagger}c_{1}c_{2}^{\dagger}c_{2}- \tilde{\mu}\left(n\right)(c_{1}^{\dagger}c_{1}+c_{2}^{\dagger}c_{2})+ \Delta(c_{1}c_{2}+c_{2}^{\dagger}c_{1}^{\dagger})- \tilde{t}\left(n\right)(c_{1}^{\dagger}c_{2}+c_{2}^{\dagger}c_{1})
\end{equation}
For the large detuning regime, we can use the following expression for the effective interaction:
\begin{equation}
    \tilde{U}\left(n\right) = U-2n\omega_{c}+\left(n+1\right)\frac{2\omega_{c}\tilde{\kappa}_{n+1}^{2}}{\omega_{c}^{2}-\tilde{\omega}_{n+1}^{2}}-n\frac{2\omega_{c}\tilde{\kappa}_{n}^{2}}{\omega_{c}^{2}-\tilde{\omega}_{n}^{2}}
\end{equation}
Note  that the original interaction $U$ can be tuned not only in strength, but also in sign by changing the number of photons in the cavity $n$, the frequency $\omega_c$ and the light-matter coupling $g$.
As a corollary, we plot in Fig.~\ref{fig:EffectiveU3} a density plot that characterizes the sign and strength of the effective interaction as a function of the cavity parameters. As predicted, for the empty cavity $n=0$ and large detuning $\omega_c$, the cavity-mediated interaction can only be repulsive. 
Instead, for a cavity with one photon $n=1$, the effective interaction can become attractive in the large detuning regime.
\begin{figure}
    \centering
    \includegraphics[width=0.45\textwidth]{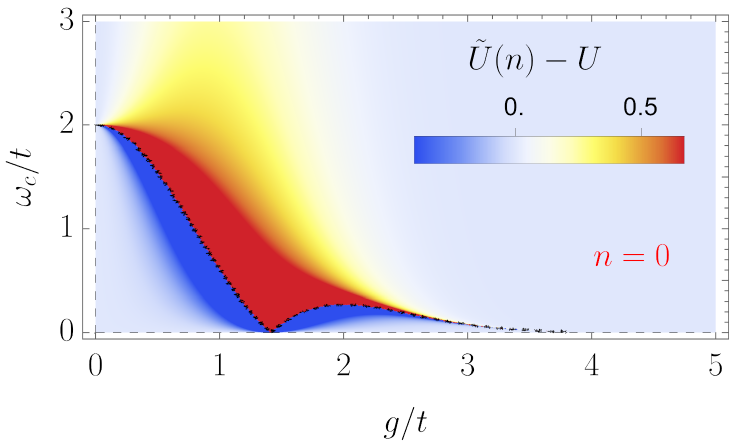}
    \includegraphics[width=0.45\textwidth]{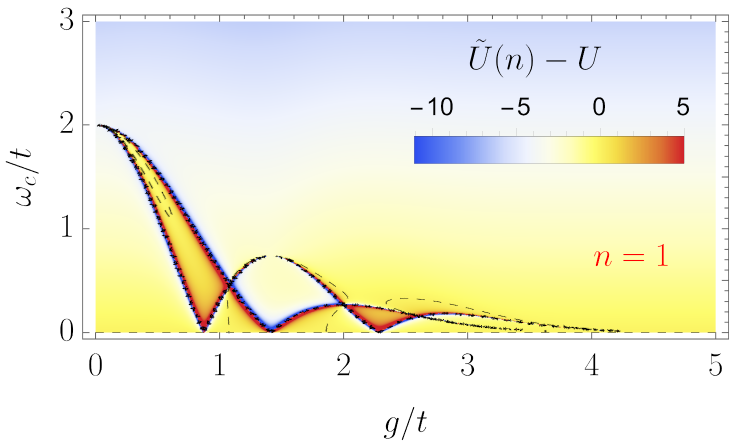}
    \caption{Cavity-mediated interaction for the $n=0,1$ subspaces (left and right, respectively) as a function of $\omega_c=$ and $g$. The black dashed lines show the values where the interaction changes sign.}
    \label{fig:EffectiveU3}
\end{figure}
These results demonstrate that the effective interaction in the system can be highly controlled via the cavity. The other effective parameters in the large detuning regime are given by
\begin{align}
    \tilde{\mu}\left(n\right) =& \mu-n\omega_{c}+\frac{\left(n+1\right)\omega_{c}\tilde{\kappa}_{n+1}^{2}}{\omega_{c}^{2}-\tilde{\omega}_{n+1}^{2}}-\frac{n\omega_{c}\tilde{\kappa}_{n}^{2}}{\omega_{c}^{2}-\tilde{\omega}_{n}^{2}} ,\\
    \tilde{t}\left(n\right) =& te^{-\frac{g^{2}}{2}}L_{n}(g^{2})-\frac{(n+1)\tilde{\omega}_{n+1}\tilde{\kappa}_{n+1}^{2}}{\omega_{c}^{2}-\tilde{\omega}_{n+1}^{2}}-\frac{n\tilde{\omega}_{n}\tilde{\kappa}_{n}^{2}}{\omega_{c}^{2}-\tilde{\omega}_{n}^{2}}
\end{align}
\section{Hamiltonian in the Majorana basis}
The original Hamiltonian can be expressed in the Majorana basis to explicitly show how MBS are present if the necessary conditions are fulfilled. Starting from the general expression:
\begin{equation}
    H=\omega_c a^{\dagger}a-\mu\left(c_{1}^{\dagger}c_{1}+c_{2}^{\dagger}c_{2}\right)-te^{-ig\left(a^{\dagger}+a\right)}c_{1}^{\dagger}c_{2}-te^{ig\left(a^{\dagger}+a\right)}c_{2}^{\dagger}c_{1}+\Delta \left( c_{1}c_{2}+ c_{2}^{\dagger}c_{1}^{\dagger} \right)+Uc^\dagger_1 c_1 c^\dagger_2 c_2
\end{equation}
we apply the transformation $c_{j}=\frac{1}{2}\left(\gamma_{2j-1}+i\gamma_{2j}\right)$ with $\left\{ \gamma_{j},\gamma_{k}\right\} =2\delta_{j,k}$. Then, after rewriting Peierls phase in terms of bosonic Hubbard operators we find that the Hamiltonian can be expressed as:
\begin{align}
    H=&\frac{U}{4}-\mu+\omega_c\sum_{n=0}^{\infty}nY^{n,n}-\frac{U}{4}\gamma_{1}\gamma_{2}\gamma_{3}\gamma_{4} \nonumber \\
    &+\frac{i}{2}\left(\frac{U}{2}-\mu\right)\left(\gamma_{1}\gamma_{2}+\gamma_{3}\gamma_{4}\right)+\frac{i\Delta}{2}\left(\gamma_{1}\gamma_{4}+\gamma_{2}\gamma_{3}\right) \nonumber \\
    &-\frac{i}{4}\sum_{n,m=0}^{\infty}Y^{m,n}\left[t_{m,n}\left(g\right)+t_{m,n}\left(-g\right)\right]\left(\gamma_{1}\gamma_{4}-\gamma_{2}\gamma_{3}\right) \nonumber \\
    &+\frac{1}{4}\sum_{n,m=0}^{\infty}Y^{m,n}\left[t_{m,n}\left(g\right)-t_{m,n}\left(-g\right)\right]\left(\gamma_{1}\gamma_{3}+\gamma_{2}\gamma_{4}\right)
\end{align}
If we first ignore the interaction $U$ and the coupling with the cavity, we find that the Hamiltonian reduces to:
\begin{equation}
    H = -\mu-\frac{i\mu}{2}\left(\gamma_{1}\gamma_{2}+\gamma_{3}\gamma_{4}\right)+\frac{i}{2}\left[\left(t+\Delta\right)\gamma_{2}\gamma_{3}-\left(t-\Delta\right)\gamma_{1}\gamma_{4}\right]
\end{equation}
Hence, if we set $\mu = 0$ and $t=\Delta$, the Hamiltonian simplifies to $H= it\gamma_{2}\gamma_{3}$ and the spectrum has two isolated MBS, $\gamma_1$ and $\gamma_4$, which have zero energy and are decoupled from the rest.
This is the standard situation for the minimal Kitaev chain and defines the well known sweet spot in the non-interacting case.

Instead, if the Hamiltonian includes interactions
\begin{equation}
    H  = \frac{U}{4}-\mu-\frac{i}{2}\left(\mu-\frac{U}{2}\right)\left(\gamma_{1}\gamma_{2}+\gamma_{3}\gamma_{4}\right) \label{eq:MajoranaH0SM}+ i\frac{\Delta-t}{2}\gamma_{1}\gamma_{4}+i\frac{\Delta+t}{2}\gamma_{2}\gamma_{3}-\frac{U}{4}\gamma_{1}\gamma_{2}\gamma_{3}\gamma_{4} ,
\end{equation}
the MBS hybridize and their localization is affected, because even for $\mu=U/2$ and $t=\Delta$, the term proportional to $U\gamma_1 \gamma_2 \gamma_3 \gamma_4$ remains.
On top of this, if the cavity also couples to the system, not only the hopping is renormalized, but also effective interactions will be induced and create a similar many-body effect. However, this can be beneficial because we can try to compensate the interaction term with the cavity-induced effective interaction.
The full Hamiltonian in the large detuning regime and the Majorana basis reads:
\begin{equation}
    \tilde{H}\left(n\right)= \frac{\tilde{U}\left(n\right)}{4}-\tilde{\mu}\left(n\right)-\frac{\tilde{U}\left(n\right)}{4}\gamma_{1}\gamma_{2}\gamma_{3}\gamma_{4}- \frac{i}{2}\left[\tilde{\mu}\left(n\right)-\frac{\tilde{U}\left(n\right)}{2}\right]\left(\gamma_{1}\gamma_{2}+\gamma_{3}\gamma_{4}\right)+ i\frac{\Delta-\tilde{t}\left(n\right)}{2}\gamma_{1}\gamma_{4}+i\frac{\Delta+\tilde{t}\left(n\right)}{2}\gamma_{2}\gamma_{3}
\end{equation}
Therefore, if we impose $\Delta=\tilde{t}(n)$, $\tilde{U}(n)=0$ and $\tilde{\mu}=\tilde{U}(n)/2$, we can cancel all terms hybridizing the MBS at the edges. In particular, the third condition can be simplified to $\mu=U/2$, which interestingly coincides with the requirement to have Majorana polarization equal to one. The remaining two conditions are equivalent to require the degeneracy between the even and odd eigenstates for both, the ground and the excited states, and can be used to determine the experimental values $\omega_c$ and $t$ that allow us to reach the new sweet spot of this c-QED setup.

In Fig.~\ref{fig:Spectrum0SM} we compare the spectrum of the isolated two-site Kitaev chain (blue dot-dashed) with the case coupled to a cavity for the large detuning regime $\omega_c\gg t,\Delta$ (red solid and black dashed line for the exact result and its analytical approximation from Eq.~\eqref{eq:Heff}, respectively).
In the isolated case, due to the interaction $U=-0.2\Delta$, the spectrum becomes highly asymmetric around zero energy. For this reason, when the even and odd ground states are degenerate, the resulting MBSs are still hybridized.
This asymmetry is a direct consequence of interactions and cannot be removed by tuning the single particle terms in the Hamiltonian of Eq.~\eqref{eq:MajoranaH0}.

When the system couples to the cavity the energy levels of the even subspace are unaffected because they do not couple to the photons. However, the energy levels of the odd subspace are modified in such a way that at the sweet spot $t_s$ (vertical dashed line), the even and odd parity subspaces become simultaneously degenerate for both, the ground and the excited state, restoring the symmetry of the spectrum around $E=0$.
The analytical approximation (black dashed lines) agrees well with exact diagonalization (red solid lines) up to large values of $t/\Delta$ when the large detuning regime is not valid anymore since the bandwidth becomes comparable to the cavity frequency.

\begin{figure}
    \centering
    \includegraphics[width=0.45\columnwidth]{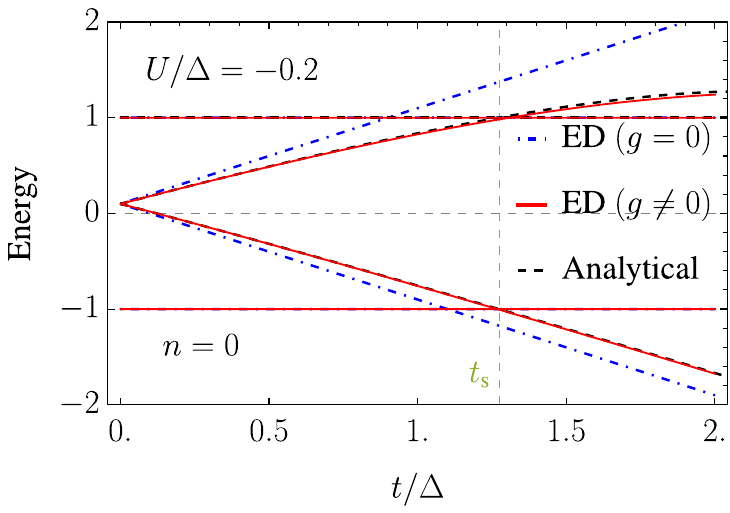}
    \caption{Spectrum of the interacting two-site Kitaev chain coupled to a cavity in its ground state. Blue dot-dashed line corresponds to the isolated chain case ($g$ = 0) given by $E_\pm^{\text{odd}}$ and $E_\pm^{\text{even}}$.  Red solid and black dashed lines correspond to exact diagonalization for $N_{max}=20$ and its analytical  approximation from Eq.~\eqref{eq:Heff}, respectively. The vertical green dashed line indicates the  sweet spot $t_\text{s} = 1.27\Delta$, obtained from  solving Eqs.~\eqref{eq:EffU0} and \eqref{eq:tEff2}. The other parameters are $\omega_c=5\Delta$, $g = 0.65$, $U=-0.2\Delta$ and $\mu=U/2$.
    }
    \label{fig:Spectrum0SM}
\end{figure}

\section{Cavity transmission}
The cavity transmission probably is the simplest method to indirectly obtain information about the system placed in the cavity. Mathematically, it is closely related to the photon propagator $D(t)=-i\theta(t)\langle[a(t),a^\dagger]\rangle$, which can be calculated with exact diagonalization or by different approximate methods. For our purposes, it will be enough to consider the Hamiltonian in each of the Fock subspaces. Hence, we will ignore photon transitions between different Fock subspaces that will only contribute with small corrections in our results. The relevant physics comes from the fact that the even parity subspace does not couple to the cavity photons, hence if the system is in the even parity ground state, the DOS shows a single resonance at the bare cavity frequency $\omega_c$. Instead, when the system is in the odd parity ground state the resonator frequency is shifted. 

To characterize the shift we use that the lowest order Hamiltonian for each Fock subspace can be written in the following form:
\begin{equation}
    H_{0}=\sum_{l,\mu}E_{\mu}(l)X^{\mu,\mu}Y^{l,l} \label{eq:H0}
\end{equation}
with $X^{\mu,\mu}$ a projector onto the eigenstate $|\mu\rangle$ of the matter Hamiltonian and $Y^{l,l}$ a projector onto the $l$ number of photons subspace.
The next step is to express the photon propagator in this basis (note that the photon operators only act on the photonic subspace and that we assume that the density matrix for the Kitaev chain is in the odd parity subspace):
\begin{align}
    D\left(\omega\right) =& \sum_{n,m}\frac{\langle n|a|m\rangle\langle m|\left[a^{\dagger},\rho_{\mu,l} \right]|n\rangle}{\omega+E_{\mu}(n)-E_{\mu}(m)+i\epsilon} \nonumber \\
    =& \frac{l+1}{\omega+E_{\mu}\left(l\right)-E_{\mu}\left(l+1\right)+i\epsilon}-\frac{l}{\omega+E_{\mu}\left(l-1\right)-E_{\mu}\left(l\right)+i\epsilon}
\end{align}
with $\rho_{\mu,l}$ the density matrix of the system, prepared in a state with a number of photons $| l \rangle$ and in the odd parity ground state of the Kitaev chain $|\mu\rangle$. Therefore, if we particularize for the case of a cavity with $n=0$, we find:
\begin{equation}
    D(\omega) = \frac{1}{\omega-\omega_{c}-te^{-\frac{g^{2}}{2}}g^{2}+i\epsilon}
\end{equation}
which shows that as soon as the ground state parity changes to odd, the cavity frequency shifts towards $\omega=\omega_c+te^{-\frac{g^{2}}{2}}g^{2}$. This is shown in Fig.~\ref{fig:DOS1}, where we can see that tuning the hopping, at the sweet spot condition $t_s$ the resonance shifts towards the predicted value due to the change in parity.
\begin{figure}
    \centering
    \includegraphics[width=0.4\textwidth]{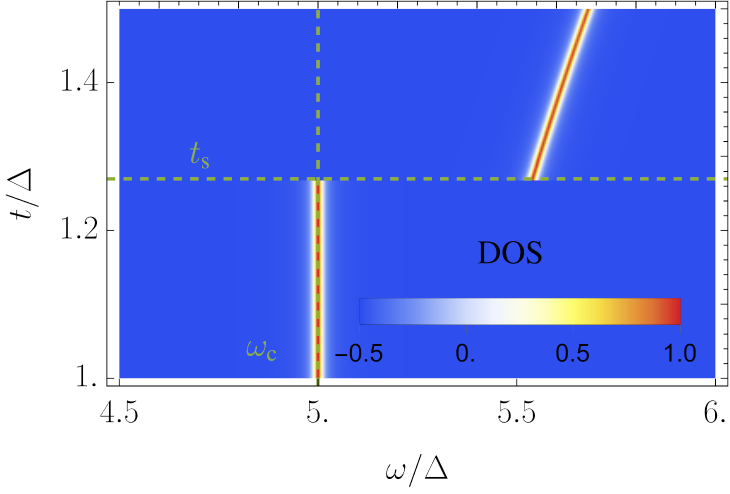}
    \caption{Density of states from the exact photon propagator for different values of the hopping $t/\Delta$ and a cavity with $n=0$ photons. Below the degeneracy point $t_{s}$, the ground state is of even parity and the response is identical to that of the unperturbed cavity. Instead, after the sweet spot value the peak shifts towards larger values due to the change of parity in the ground state. Parameters: $\omega_c=5\Delta$, $\mu=U/2=-0.1\Delta$ and $g=0.65$.}
    \label{fig:DOS1}
\end{figure}
If the cavity is in a different photon number subspace, the frequency shift changes accordingly and for example, in the case $n=1$ the photon propagator reduces to
\begin{equation}
    D(\omega) = \frac{2}{\omega-\omega_{c}-te^{-\frac{g^{2}}{2}}g^{2}\left(1-\frac{g^{2}}{2}\right)+i\epsilon}-\frac{1}{\omega-\omega_{c}-te^{-\frac{g^{2}}{2}}g^{2}+i\epsilon}
\end{equation}
The interesting feature in this case is that two peaks simultaneously emerge for $n>0$, with different sign and weight. This allows to detect the sweet spot tunneling $t_s$, but noticing that the distance between these peaks only depends on the light-matter coupling $g$, this also provides a way to measure it from the DOS. The corresponding plot for this situation is shown in the main text, demonstrating that even in situations where the intrinsic interactions are very large and the degeneracy point is impossible to reach, coupling the chain to a cavity with $n>0$ provides a way to change the parity of the ground state.
\section{Majorana polarization}

Majorana polarization~\cite{Aksenov2020MP} has been introduced as a way to quantify the quality of the MBS in more complex QD-based systems that include electron-electron interactions~\cite{tsintzis2022creating}. Defining the Majorana polarization as~\cite{souto2024subgap}
\begin{align}
    M_j = \dfrac{\langle o|c_j+c_j^\dag|e\rangle^2 - \langle o|c_j -c_j^\dag|e\rangle^2}{\langle o|c_j+c_j^\dag|e\rangle^2 + \langle o|c_j -c_j^\dag|e\rangle^2},
    \label{eq:MajoranaPolarization}
\end{align}
where $|e\rangle$ ($|o\rangle$) is the even (odd) ground state, we calculate $M_j$ in a two-site Kitaev chain coupled to cavity described by the effective Hamiltonian~\eqref{eq:Heff}. We find that the eigenstates of~\eqref{eq:Heff} are independent of the light-matter coupling strength $g$ and cavity frequency $\omega_c$ (and are the same for the cavity with $n = 0$ and $n=1$) giving rise to the Majorana polarization 
\begin{align}
    M_{j} = \dfrac{2 \Delta }{\sqrt{4 \Delta ^2+(U-2 \mu )^2}}.
    \label{eq:MPn=0}
\end{align}

 The Majorana polarization in a two-site Kitaev chain coupled to cavity~\eqref{eq:MPn=0} is equal to the one obtained for the isolated Kitaev chain~\cite{souto2023probing}. This demonstrates that the MBS arising in the two-site Kitaev chain coupled to cavity \eqref{eq:Heff} are similar to those of the isolated chain. Moreover, we note that for $\mu = U/2$ the Majorana polarization is always equal to $1$.

\bibliography{bibliographyDraftKitaev}
\end{widetext}

\end{document}